\documentstyle[12pt]{article}

\begin{document}

\font\fifteen=cmbx10 at 15pt
\font\twelve=cmbx10 at 12pt

\def\be{\begin{equation}}
\def\ee{\end{equation}}

\begin{titlepage}

\begin{center}

\renewcommand{\thefootnote}{\fnsymbol{footnote}}

{\twelve Centre de Physique Th\'eorique
\footnote{
Unit\'e Propre de Recherche 7061
}, CNRS Luminy, Case 907}

{\twelve F-13288 Marseille -- Cedex 9}

\vspace{1 cm}

{\fifteen MELOSH ROTATION AND \\
THE NUCLEON TENSOR CHARGE}

\vspace{0.3 cm}

\setcounter{footnote}{0}
\renewcommand{\thefootnote}{\arabic{footnote}}

{\bf
Ivan SCHMIDT\footnote{
Departamento de F\'\i sica, Universidad T\'ecnica Federico Santa
Mar\'\i a, Casilla 110-V \\ \null\quad Valpara\'\i so, Chile - Email
ischmidt@fis.utfsm.cl } and Jacques SOFFER\footnote{Email
soffer@cpt.univ-mrs.fr}  }

\vspace{2,3 cm}

{\bf Abstract}

\end{center}

By making use of the effect of the Melosh rotation, we show that one
can estimate, in a simple way, the nucleon tensor charge in a
relativistic quark model formulated on the light-cone. We discuss
the physical significance of our results and compare them with
those recently obtained in different phenomenological models.

\vspace{2 cm}


\bigskip

\noindent March 1997

\noindent CPT-97/P.3460

\bigskip

\noindent anonymous ftp or gopher: cpt.univ-mrs.fr

\renewcommand{\thefootnote}{\fnsymbol{footnote}}

\end{titlepage}

\setcounter{footnote}{0}
\renewcommand{\thefootnote}{\arabic{footnote}}

%\section{Introduction}

In high-energy processes, the nucleon structure is described by a
set of parton distributions, some of which are fairly well known and
best measured in deep inelastic scattering (DIS). In particular from
unpolarized DIS, one extracts the quark distributions $q(x)$, for
different flavors $q=u,d,s, etc...$, which are related to the
forward nucleon matrix elements of the corresponding {\it vector}
quark currents $\bar q\gamma^{\mu}q$, and likewise for antiquarks.
Similarly from longitudinaly polarized DIS, one obtains the quark
helicity distributions $\Delta q(x)=q_+(x)-q_-(x)$, where $q_+(x)$
and $q_-(x)$ are the quark distributions with helicity parallel and
antiparallel to the proton helicity. Clearly the spin-independent
quark distribution $q(x)$ is $q(x)=q_+(x)+q_-(x)$. We recall that
for each flavor the {\it axial charge} is defined as the first
moment of $\Delta q(x) + \Delta\bar q(x)$ namely,

\begin{equation}
\Delta q= \int_{0}^{1} dx\left[ \Delta q(x) + \Delta \bar q
(x)\right]
\end{equation}
and in terms of the matrix elements of the {\it axial} quark current
$\bar q\gamma^{\mu}\gamma^5q$, it can be written in the form

\be\label{Delta}
2\Delta qs^{\mu} = <p,s|\bar q\gamma^{\mu}\gamma^5 q|p,s>\ ,
\ee
where $p$ is the nucleon four-momentum and $s_{\mu}$ its
polarization vector. In addition to $q(x)$ and $\Delta q(x)$, for
each quark flavor, there is another spin-dependent distribution,
called the transversity distribution $h_1^q(x)$ related to the
matrix elements of the {\it tensor} quark current $\bar
q\sigma^{\mu\nu}i\gamma^5q$. The $h_1$ distribution measures the
difference of the number of quarks with transverse polarization
parallel and antiparallel to the proton
transverse polarization. One also defines the {\it tensor charge} as
the first moment
\be
\delta q=\int_{0}^{1} dx\left[ h_1^q (x) - h_1^{\bar q}(x)\right]\ .
\ee

The existence of $h_1^q(x)$ was first observed in a systematic study
of the Drell-Yan process with polarized beams {\cite{RS}} and some of
its relevant properties were discussed later in various
papers {\cite{AM,CPR,JJ}}. We recall that $q(x)$, $\Delta q(x)$ and
$h_1^q(x)$, which are of fundamental importance for our
understanding of the nucleon structure, are all leading-twist
distributions, but one can also consider a more complete set
including several higher-twist distributions {\cite{JJ}. Due to
scaling violations, these quark distributions have a
$Q^2$-dependence governed by QCD evolution equations, which are different
in the three cases. On the experimental side, a vast programme of
measurements in unpolarized DIS has been going on for more than twenty
five years, and has yielded an accurate determination of the $x$ and
$Q^2$ dependence of $q$ (and $\bar q$) for various flavors. From
several fixed-targets experiments operating now at CERN, SLAC and
DESY, we also begin having a good insight into the different quark
helicity distributions $\Delta q$, although the present available
$x$ and $Q^2$ ranges are rather limited. Concerning $h_1^q$ (or
$h_1^{\bar q}$), they are not simply accessible in DIS because they
are in fact chiral-odd distributions and they can be best extracted
in polarized Drell-Yan processes or in $Z$ production with two transversely
polarized proton beams. Such experiments will be undertaken with the
polarized $pp$ collider at RHIC \cite{M}, but so far we have no
experimental information on the shape and magnitude of these quark
transversity distributions. However there are several different theoretical
determinations of $h_1^q$ using either the MIT bag model \cite{JJ}
or QCD sum rules \cite{IK}, and also based on either a chiral
chromodielectric model \cite{BCD} or a chiral quark-soliton model
\cite{KPG,PP}.

At this point let's mention the following positivity constraint
\be
q(x) + \Delta q(x) \geq 2|h_1^q(x)|\ ,
\ee
which is exact in the parton model and valid for each flavor,
likewise for antiquarks \cite{S}.
Although some doubts have been expressed on its validity in
perturbative QCD \cite{GJJ}, it has been recently demonstrated, from
the two loop $Q^2$ evolution \cite{V}, that if the inequality is
satisfied
for a certain value of $Q^2$, it remains valid for higher $Q^2$.
As we will see, this inequality is very useful, given the present
poor knowledge we have on $h_1^q$.

In the non-relativistic quark model, transversely polarized quarks
are in transverse spin states, which by rotational invariance
implies that the axial charge and the tensor charge must be equal.
For example by using the $SU(6)$ proton wavefunction one finds the
well known valence contributions
\be
\Delta u = \delta u = 4/3\ ,\ \Delta d = \delta d = -1/3 \ \mbox{ and
}\ \Delta s = \delta s = 0\ .
\ee
So in this case the sum of spin of quarks (and antiquarks) are equal
to the proton spin at rest since from eq.~(5) we have
\be
\Delta \Sigma \equiv \Delta u + \Delta d + \Delta s = 1\ .
\ee

Of course in DIS one is probing the proton spin in the infinite
momentum frame and the above result might not be longer true. In the
light-cone formalism,which is suitable to describe the relativistic
many-body problem, we have to transform the instant quark states
$q^i_{NR,\lambda}$ into the light-cone quark states
$q^i_{LC,\lambda}  (i=1,2,3)$.
The two sets of states are related by a general Melosh rotation
\cite{ME}, according to
\be
\begin{array}{lcl}
q^i_{LC,+}&=&\frac{1}{\sqrt{det}}\left[(m_q+x_i{\cal M})q^i_{NR,+} +
k_i^R
q^i_{NR,-}\right]\ ,\cr
\cr
q^i_{LC,-}&=&\frac{1}{\sqrt{det}}\left[- k_i^L
q^i_{NR,+}+(m_q+x_i{\cal M})q^i_{NR,-} \right]\ ,\cr
\end{array}
\ee
where $k_i^{R,L}=k_i^1\pm ik_i^2$ , $det = (m_q+x_i{\cal M})^2+\vec
k_{iT}^2$ and the invariant mass ${\cal M}$ is given by ${\cal
M}^2=\sum_{i=1}^{3} (\vec k_{iT}^2 + m_q^2)/x_i$. Here we are using
light-cone momentum fractions $x_i=p_i^+/P^+$, where $P$ and $p_i$
are the nucleon and quark momenta respectively ($p_i^+= p_i^0+p_i^3$
and $P^+= P^0 +P^3$), and the internal momentum variables $\vec
k_{iT}$ are given by $\vec k_{iT} = \vec p_{iT} - x_i \vec P_T$ with
the constraints $\displaystyle \sum_{i=1}^{3} \vec k_{iT} = 0$ and
$\displaystyle \sum_{i=1}^{3} x_i = 1$. In the zero binding limit
$x_i{\cal M}\to k_i^+ = k_i^0 + k_i^3$, but this cannot be a
justified approximation for QCD bound states.

We notice that the helicity states get mixed as long as the internal
transverse momentum $k_T$ is non-zero, which makes the Melosh
rotation non-trivial. Actually, one can show that for light-cone
states only the positive component of the axial current contributes,
so eq.~(\ref{Delta}) reads also as
\be
2\Delta q_{LC} = < p,s|\bar q_{LC,\lambda} \gamma^+\gamma^5
q_{LC,\lambda}|p,s>
\ee
with $\lambda = +$ or $-$.

By using eq.~(7) one sees that the light-cone axial charge $\Delta
q_{LC}$ is related to the non-relativistic axial charge $\Delta
q_{NR}$ as follows \cite{MA}
\be
\Delta q_{LC} = <M_q>\Delta q_{NR}\ ,
\ee
where
\be
M_q = \frac{(m_q+x_3{\cal M})^2 - k_{3T}^2}{(m_q+x_3{\cal M})^2 +
k_{3T}^2}
\ee
and $<M_q>$ is its expectation value
\be
<M_q> = \int d^3k\ M_q\ |\Psi (k)|^2\ ,
\ee
where $\Psi (k)$ is a simple normalized momentum wavefunction. By
choosing two different reasonable wavefunctions, e.g. the  harmonic
oscillator and the power-law fall off, the calculation \cite{BS}
gave $<M_q> = 0.75$ ($q=u,d$ if we assume $m_u=m_d$) which leads to
a reduction of $\Delta\Sigma$ (see eq.~(6)) from $1$, in the naive
quark model, to $0.75$.
>From polarized DIS, one obtains the singlet axial charge of the
proton $a_0(Q_0^2)=0.28\pm 0.16$ at $Q_0^2=10~GeV^2$ \cite{A}, which
is related to $\Delta\Sigma$ in a scheme dependent way, and from the
value of the gluon polarization $\Delta g \sim 2$, it implies
$\Delta\Sigma\sim 0.5$. This shows that although relativistic
effects do not provide the correct result, they are responsible for
a substantial shift in the right direction.

Let us now turn to the tensor charge. Like for the axial charge, it
can be shown that for light-cone states only the positive component
of the tensor quark current, which involves a spin-flip, contributes
so we have
\be
2\delta q_{LC} = <p,s|\bar q_{LC,\lambda}\gamma^+\gamma^{\perp}
q_{LC,-\lambda}|p,s>,
\ee
with $\lambda = +$ or $-$ and $\gamma^{\perp}=\gamma^1+i\gamma^2$.
By using eq.~(7) are easily finds how the light-cone tensor charge is
related to the non-relativistic one, namely
\be
\delta q_{LC} = <\widetilde{M}_q>\delta q_{NR},
\ee
where
\be
\widetilde{M}_q = \frac{(m_q+x_3 {\cal
M})^2}{(m_q+x_3{\cal M})^2+k_{3T}^2}
\ee
and $<\widetilde{M}_q>$ is its expectation value, as before for
$M_q$. At this point we note that in the non-relativistic case,
which corresponds to the limit $k_T=0$, one has
$M_q=\widetilde{M}_q=1$ and they both decrease under the
relativistic effects. In addition it is interesting to remark that
one has
\be
1 + M_q = 2\widetilde{M}_q,
\ee
which means that there is saturation of eq.~(4). By taking the
expectation value of eq.~(15) and knowing that $<M_q>=3/4$, as
indicated above, one finds immediately $<\widetilde{M}_q>=7/8$. By
using eq.~(5) it leads to
\be
\delta u_{LC} = 4/3 \times 7/8 = 7/6 \  \mbox{ and }\  \delta d_{LC} =
-1/3\times 7/8 = -7/24\ ,
\ee
which remarkably are exactly the values obtained in the MIT bag model
\cite{HJ}. It is worth recalling that the MIT bag model produces
quark distributions which also saturate eq.~(4) and this is also the
case for the toy model proposed in ref. \cite{AM}. These values are
perfectly compatible with the positivity bounds derived in ref.
\cite{S}, namely
\be
|\delta u|\leq 3/2\  \mbox{ and }\  \ |\delta d|\leq 1/3\ .
\ee
However in ref. \cite{KPG} they obtain
\be
\delta u = 1.12 \  \mbox{ and }\  \delta d = -0.42
\ee
but the large $N_c$ behavior is expected to generate in this model,
large theoretical uncertainties, mainly for the $d$ quark.

Unlike the axial charge which is $Q^2$-independent, $\delta q$ has the
following $Q^2$ evolution \cite{AM}
\be
\delta q(Q^2) = \delta
q(Q^2_0)\left[\frac{\alpha_s(Q^2)}{\alpha_s(Q^2_0)}\right]^{\frac{4}{33-2N_f}}
\ee
where $N_f$ is the number of flavors. So $\delta q$ decreases for
increasing $Q^2$ and in the model of ref. \cite{BCD} where the
initial scale of the nucleon is taken to be $Q^2_0 = 0.16~GeV^2$,
they obtain at $Q^2=25~GeV^2$ from eq.~(19)
\be
\delta u = 0.969  \  \mbox{ and }\  \delta d = -0.250\ ,
\ee
also consistent with the positivity bounds eq.~(17).

The relativistic light-cone quark model then predicts the following
values for the nucleon's isovector and isoscalar tensor charges,
respectively
\be
\delta u - \delta d = 35/24 = 1.458 \  \mbox{ and }\  \delta u + \delta
d = 7/8 = 0.875\ .
\ee

Let us now consider the non-relativistic and the ultra-relativistic
limits in this formalism. One simple way \cite{BS} to obtain both
limits is by varying the dimensionless quantity $M_pR_1$, where
$M_p$ is the proton mass and $R_1$ is the proton radius ($R_1 =
\sqrt{-6\ dF_1(Q^2)/dQ^2\vert_{Q^2=0}}$, where $F_1 (Q^2)$ is the
Dirac form factor). The non-relativistic limit corresponds to
$R_1\to \infty$ with a fixed mass. Using eq.~(15) and the fact that
$<M_q>\to 1$ in the $NR$ limit, as it should from eq.~(14), we obtain
that in this limit $<\widetilde{M}_q>\to 1$ also, which is then
consistent. The ultra-relativistic limit is obtained by taking
$M_pR_1\to 0$, which then corresponds to a point-like particle.
Again we use eq.~(15), but now $<M_q>\to 0$ \cite{BS}, which means that
$<\widetilde{M}_q>\to 1/2$ in the $UR$ limit. Thus the point-like
values for the tensor charges are predicted to be
\be
\delta u\to 2/3\  \mbox{ and }\  \delta d\to - 1/6\ .
\ee

\subsection*{Acknowledgements}

This research has received the financial support of the scientific
cooperation programme ECOS-CONICYT between France and Chile, and also
of Fondecyt (Chile), contract 1960536. One of (J.S.) is grateful
to Universidad T\'ecnica Federico Santa Mar\'\i a in Valpara\'\i so
for warm hospitality during the elaboration of this work.


\begin{thebibliography}{99}

\bibitem{RS} J.P. RALSTON and D.E. SOPER, {\it Nucl. Phys.},  {\bf
B152} (1979) 109.

\bibitem{AM} X. ARTRU and M. MEKHFI, {\it Z. Phys.},  {\bf
C45} (1990) 669.

\bibitem{CPR} J.L. CORTES, B. PIRE and D.E. J.P. RALSTON, {\it Z.
Phys.},  {\bf C55} (1992) 409.

\bibitem{JJ} R.L. JAFFE and X.JI, {\it Phys. Rev. Lett.},  {\bf
67} (1991) 552 ; {\it Nucl. Phys.}, {\bf B375} (1992) 527.

\bibitem{M} Y. MAKDISI, Proceedings of the 12$^{th}$ Int.
Conference on High Energy Spin Physics, Amsterdam, Sept. 1996.

\bibitem{IK} B.L. IOFFE and A. KHODJAMIRIAN, {\it
Phys. Rev.},  {\bf D51} (1995) 33.

\bibitem{BCD} V. BARONE, T. CALARCO and A. DRAGO, {\it
Phys. Lett.},  {\bf B390} (1997) 287.

\bibitem{KPG} H.-C. KIM, M.V. POLYAKOV and K. GOEKE, {\it
Phys. Lett.},  {\bf B387} (1996) 577.

\bibitem{PP}P.V. POBYLITSA and M.V. POLYAKOV, {\it
Phys. Lett.},  {\bf B389} (1996) 350.

\bibitem{S} J. SOFFER, {\it Phys. Rev. Lett.},  {\bf
74} (1995) 1292.

\bibitem{GJJ} G.R. GOLDSTEIN, R.L. JAFFE and X. JI, {\it
Phys. Rev.},  {\bf D52} (1995) 5006.

\bibitem{V} W. VOGELSANG, Contribution to the Ringberg Workshop
on High Energy Polarization Phenomena, Feb. 24-28
(1997).

\bibitem{ME} H.J. MELOSH, {\it
Phys. Rev.},  {\bf D9} (1974) 1095.

\bibitem{MA} B.-Q. MA, {\it J. Phys.}, {\bf G17} (1991) L53 ; B.-Q.
MA and Q.-R. ZHANG {\it Z. Phys.},  {\bf C58} (1993) 479.

\bibitem{BS}S.J. BRODSKY and F. SCHLUMPF, {\it
Phys. Lett.},  {\bf B329} (1994) 111.

\bibitem{A} D. ADAMS et al. (Spin Muon Collaboration) hep-ex/9702005.

\bibitem{HJ} H. HE and X. JI, {\it
Phys. Rev.},  {\bf D52} (1995) 2960.



\end{thebibliography}
\end{document}